\documentclass{jkas}

\def\year{2016} 
\def\month{September} 
\def\volume{00} 
\def\issue{0} 
\def\beginpage{1} 
\def\endpage{99} 
\def\received{---} 
\def\accepted{---} 

\date{Received \received ; accepted \accepted}

\def\au{{\rm AU}}

\def\muas{{\mu\rm as}}
\def\kpc{{\rm kpc}}

\def\min{{\rm min}}
\def\mas{{\rm mas}}
\def\lim{{\rm lim}}

\def\max{{\rm max}}

\def\rel{{\rm rel}}

\def\apj{{ApJ}}
\def\aj{{AJ}}

\def\mnras{{MNRAS}}

\def\au{{\rm AU}}

\title{
     Gaia Parallax Zero Point From RR Lyrae Stars
}

\author[1,2,3]{Andrew Gould}
\author[4]{Juna A. Kollmeier}

\affil[1]{Max-Planck-Institute for Astronomy, K\"onigstuhl 17, 69117 Heidelberg, Germany; \email{gould@astronomy.ohio-state.edu }}
\affil[2]{Korea Astronomy and Space Science Institute, Daejon 305-348, Republic of Korea}
\affil[3]{Department of Astronomy Ohio State University, 140 W.\ 18th Ave., Columbus, OH 43210, USA}
\affil[4]{Observatories of the Carnegie Institution of Washington, 813 Santa Barbara Street, Pasadena, CA 91101}

\def\corrauthor{
Andrew Gould
}

\def\runningauthor{
Gould \& Kollmeier

}

\def\runningtitle{
    Gaia Parallax Zero Point From RR Lyrae Stars
}

\def\keywords{
astrometry
}

\def\abstracttext{
Like Hipparcos, Gaia is designed to give absolute parallaxes, independent of any astrophysical reference system.  And indeed, Gaia's
internal zero-point error for parallaxes is likely to be smaller than any individual parallax error.  Nevertheless, due in part to
mechanical issues of unknown origin, there are many astrophysical questions for which the parallax zero-point error $\sigma(\pi_0)$ will be the
fundamentally limiting constraint.  These include the distance to the Large Magellanic Cloud and the Galactic Center.  We show that by using the photometric parallax estimates for RR Lyrae stars (RRL) within 8kpc, via the ultra-precise infrared period-luminosity relation, one
can independently determine a hyper-precise value for $\pi_{0}$. Despite their paucity relative to bright quasars, we show that RRL are
competitive due to their order-of-magnitude improved parallax precision for each individual object relative to bright quasars. We
show that this method is mathematically robust and well-approximated by analytic formulae over a wide range of relevant distances.
}

\begin{document}
\jkashead 


\section{Introduction \label{sec:intro}}

Gaia will obtain astrometry for $>10^9$ stars, with parallax precisions
down to $\sigma(\pi)\lesssim 6\muas$ for the brightest stars,
$V\lesssim 12$.  In contrast to traditional pre-Hipparcos astrometry,
Gaia is designed to measure so-called ``absolute parallaxes''.
However, since nothing in nature is truly ``absolute'',
it behooves us to specify more precisely exactly what Gaia will
measure.

In traditional narrow angle astrometry, one measures the parallactic
motion of some target star relative to a set of reference stars,
and from this measures the ``relative parallax'' 
$\pi_\rel = \pi_{\rm target} - \pi_{\rm reference}$, where the last
quantity is the mean parallax of the reference stars.  One then
estimates the distances of the reference stars, and hence $\pi_{\rm reference}$,
by some non-astrometric method, usually photometric. If, for example,
the reference stars are five times farther than the target star,
and if their distances can be estimated to, say 30\% precision, 
and assuming $N=4$ reference stars, then
the contribution of the error due to the reference frame is only
$\sigma(\pi_{\rm reference})/\pi_{\rm target} = 30\%/5/\sqrt{4}=3\%$,
which may well be lower than the contribution from the astrometric
precision of the $\pi_\rel$ measurement.  In a variant of this approach,
one might use external quasars or galaxies as the reference frame, in which
case $\pi_{\rm reference}=0$ to a precision adequate for most purposes.

By contrast Hipparcos used wide-angle astrometry, which does not
require any external reference frame for parallaxes (although it
does for proper motions).  To understand the basic principle of this
approach, consider two telescopes that are rigidly separated by $90^\circ$.
Let the first of these telescopes make two measurements of a star
in the ecliptic,
six months apart, both times at quadrature.  That is, both of these
measurements will suffer maximal parallactic deflection, but in
opposite directions.  Now let the second telescope measure the positions
of a second star, also in the ecliptic at the same epochs.  Since this second
star is, by construction, aligned perfectly with the Sun, it will not suffer any parallactic
deflection.  Hence the relative change in position of these two stars
directly gives the absolute parallax of the first.  Now, of course,
one of these two measurements of the second star could not be made
in practice because it would lie directly behind the Sun.  However,
the point is that by simultaneously observing stars that are
affected by parallax by substantially different (and easily calculable)
amounts, one can extract the absolute parallax.

For this method to work to a given specified precision, the
``basic angle'' between the two telescopes must remain fixed to
the same precision.  Or rather, any changes in the basic angle
must be understood to this specified level of precision.  If the
basic-angle oscillations have power on timescales shorter than
the rotation period of the two telescopes, then the amplitude
of these oscillations can be derived (and so corrected)
from the observations themselves.  However, oscillations at
the rotation period are indistinguishable from a zero-point
offset of all parallaxes that are being measured.  Uncertainty
about this amplitude is therefore equivalent to introducing a
``$\pi_{\rm reference}$'' term, as in narrow-angle astrometry.

For reasons that are not presently understood, the actual
amplitude of these oscillations is about 1 mas, which is
orders of magnitude higher than expected from the original
design, and also orders of magnitude higher than the parallax
precision of the best measurements.  Happily, the great majority
of this oscillation can be measured from engineering data,
but an ultra-precise estimate of the Gaia system parallax zero
point, $\pi_0$, will require external calibration.

It may well be that for most applications a precision determination
of $\pi_0$ is irrelevant.  However, it is easy to imagine applications
for which this is important.  For example, the parallax of the 
Large Magellanic Cloud (LMC)
is presently estimated to be $\pi_{\rm LMC}=20\,\muas$ (for a comparison of measurements see \citealt{degrijs14}).  Consider measurements
of 10,000 LMC stars at $V=16$, each with precision $\sigma(\pi)=40\,\muas$.
Each measurement by itself would be ``useless'', having a 200\% error.
Nevertheless, the combination of all of them would have an error
$\sigma(\pi_{\rm LMC})=0.4\,\muas$, i.e., a 2\% error.  However, if the
zero-point error $\sigma(\pi_0)\sim 2\,\muas$, then this LMC distance
measurement would be degraded by a factor 5.

One method to measure $\pi_0$ is from quasars.  There is about
one such object per square degree to $V_0=18$ (e.g., \citealt{hewett01}).  For the $\sim 3/4$
of these that are relatively unextincted, the Gaia 
precision\footnote{The Gaia site http://www.cosmos.esa.int/web/gaia/science-performance gives $\sigma(\pi) =(-1.631 + 680.766 · z + 32.732 · z2)1/2 · [0.986 + (1 - 0.986) · (V-I)$ where $z=\min(10^{0.4(G-15)},10^{-1.2})$}
is anticipated to be $\sigma(\pi)\sim 140\,\muas$.  Since these are
each known a priori to have zero parallax (or rather $\pi\ll 1\,\muas$),
the 30,000 that lie over $3\pi$ sterradians can be combined to yield
$\sigma(\pi_0)\sim 0.8\,\muas$.   There are $\sim 3$ times more
quasars $(18<V_0<19)$ than $V_0<18$,  but each contributes substantially
less information.  Including all quasars, we estimate 
$\sigma(\pi_0)\sim 0.6\,\muas$ from this technique.  

This estimate then sets the benchmark for other techniques.  If these
other methods can achieve a similar or better precision, then they
can serve as an independent check on the quasars and improve the
overall measurement of $\pi_0$.

\section{{RR Lyrae Star Based Zero Point: Naive ``Circular'' Argument}
\label{sec:naive}}

At infrared wavelengths, RR Lyrae stars obey a period-luminosity (PL)
relation
\begin{equation}
L_\lambda = L_{0,\lambda}(P/P_0)^\beta
\label{ref:pl}
\end{equation}
where $P_0$ is chosen to be near the mean period of the sample (\citealt{longmore86}, \citealt{longmore90}).
There is some scatter around this relation, which is usually
expressed in magnitudes 
$\sigma(M_\lambda) = (5/\ln 10)\langle(\delta L)^2\rangle^{1/2} /L$,
but which we will express for convenience in terms of the error
in inferred distance
\begin{equation}
\epsilon = {\langle(\delta L)^2\rangle^{1/2}\over L}
\label{ref:epsdef}
\end{equation}
At present, $\epsilon$ is not known because it appears to be
below the precision of the best RR Lyrae parallax measurements
made to date (e.g. \citealt{benedict11, madore13, dambis14, braga15}).  It may plausibly be  $\epsilon\sim 0.01$ or even less depending on wavelength (although see theoretical estimates from \citealt{bono01}).  This scatter will 
be easily probed by Gaia.  RR Lyrae stars with distance $D<2\,\kpc$
will have parallax errors $\sigma(\pi)\sim 6\,\muas$ and therefore
fractional distance errors $\sigma(\pi)/\pi\sim 0.6\% (D/1\,\kpc)^{-1}$.
Thus, it will be quite noticeable if the nearby RR Lyrae stars
show intrinsic luminosity scatter such that $\epsilon>0.01$.

The basic approach then is to measure $L_0$ using the relatively
nearby stars $D\lesssim 2\,\kpc$ (see recent overview by Beaton et al. 2016).  Because their parallaxes are so much larger than any possible zero point error, the latter can to first
approximation be ignored, and the very high precision parallax measurements can then be used to measure $L_0$ (as well as $\sigma$).  Then one can apply this
knowledge to much more distant RR Lyrae stars, e.g., at $D\sim 5\,\kpc$.
While the parallaxes of these stars $(\pi\sim 200\,\muas)$ are also much
larger than any possible $\pi_0$, the error in the individual
distances due to scatter in the PL relation is only $\epsilon\pi\sim 2\,\muas$.
Not only is this now of order the plausible values of $\pi_0$, more to
the point it is much smaller than the Gaia measurement error for these
stars, $\sigma(\pi)\sim 15\,\muas$.   Since the uncertainty in the
photometric parallax estimate is much smaller than the parallax
measurement error, it basically does not contribute.   Hence, each
parallax measurement of such relatively distant RR Lyrae stars
constitutes an independent estimate of $\pi_0$ with error $15\,\muas$ per star.
Thus, even though there are many fewer RR Lyrae stars than quasars,
they can be competitive because of much smaller errors for each
measurement.

Then, with $\pi_0$ measured, one can go back and improve the
determination of $L_0$ by properly accounting for this zero-point
offset.  The last step may appear circular, but we will
see that each element of this description, including the naively
``circular'' argument, maps directly onto a rigorous statistical
approach.

\section{{Mathematical Description}
\label{sec:math}}

Strictly speaking we should simultaneously fit for four parameters,
$L_0$, $\beta$, $\epsilon$, and $\pi_0$.  However, $\beta$ and
$\epsilon$ are essentially uncorrelated from the other parameters.
In the interests of focusing on the main determinants of the problem,
we will take $\beta$ and $\epsilon$ as given.

We can then write the relation between observed and modeled
parallaxes
\begin{equation}
\pi_{\rm obs,k} = \pi_0 + A\pi_{{\rm fid},k}\pm \sigma_k;
\quad
\pi_{\rm fid}\equiv \sqrt{4\pi F_{{\rm dered},k}\over L_{0,\rm fid} (P/P_0)^\beta}
\label{eqn:model}
\end{equation}
where $F_{{\rm dered},k}$ is the dereddened observed flux of the $k$th
star in the appropriate infrared band, and $L_{0,\rm fid}$ is the
initial guess for $L_0$ (which will then be corrected by measuring $A$).
The error $\sigma_k$ is the quadrature sum of two contributions.
The first is from the scatter in the PL relation, namely $\epsilon\pi$.
The second is the measurement error.  For now we will assume that
all the stars are in the photon limit and that therefore this error
is inversely proportional to the square root of the flux in the Gaia
bands.  Since RR Lyrae luminosities are roughly independent
of period in optical bands, this implies (if we restrict attention
to relatively unextincted stars), that the error is inversely proportional
to the flux.  Assuming that $M_G=0.6$ in the Gaia band, and adopting
the anticipated Gaia precision in the photon limit, one then finds
\begin{equation}
\sigma^2(\pi) = (\epsilon\pi)^2 + \biggl({\kappa\over\pi}\biggr)^2
\qquad \kappa = (57\,\muas)^2
\label{eqn:sigma}
\end{equation}

We then follow the standard procedure of constructing a Fisher matrix
and approximating it as an integral
(e.g., \citealt{gould95}).  First, one forms the inverse
covariance matrix of the two parameters $(\pi_0,A)$, which are
labeled ``0'' and ``1'', respectively 
\begin{equation}
B_{ij} = \sum_k {\pi_k^{i+j}\over \epsilon^2 \pi_k^2 + \kappa^2/\pi_k^2}
= {1\over\epsilon^2}\sum_k {\pi_k^{i+j-2}\over 1 + (\kappa/\epsilon)^2/\pi_k^4}
\label{eqn:bij}
\end{equation}
Switching variables to distance $r=\au/\pi$
and taking the sum to an integral, we obtain
\begin{equation}
B_{ij} 
= {1\over\epsilon^2}\sum_k {(r_k/\au)^{2-i-j}\over 1 + (r_k/D_*)^4}
\label{eqn:bij2}
\end{equation}
\begin{equation}
B_{ij} 
\rightarrow {3\pi n(\au)^{i+j-2}\over \epsilon^2}\int_0^{r_\max} dr\,
{r^{4-i-j}\over 1 + (r/D_*)^4}
\label{eqn:bij3}
\end{equation}
where we have assumed a uniform density $n$ and that the RR Lyrae stars 
can be effectively incorporated only over $3\pi$ sterradians.
Here 
\begin{equation}
D_*\equiv\au\sqrt{\epsilon\over\kappa}.
\label{eqn:ddef}
\end{equation}
Substituting $x=r/D_*$ yields
\begin{equation}
B_{ij} 
= {3\pi n D_*^3\over \epsilon\kappa}
\biggl({\kappa\over\epsilon}\biggr)^{i+j\over 2}b_{ij}(x_\max),
\quad
b_{ij}(x) = 
\int_0^x dy
{y^{4-i-j}\over 1 + y^4}
\label{eqn:bij4}
\end{equation}

Unfortunately, only the off-diagonal terms of $b_{ij}$ can be evaluated
in closed form, $b_{01}(x)=\ln(1+x^4)^{1/4}$.  However, for $x\gtrsim 2.5$,
$b$ very quickly approaches its asymptotic 
limit\footnote{This is because the next terms in the expansion
are $+x^{i+j-5}/(5-i-j)$, i.e., 4 powers of $x$ below the leading term
(plus an additional factor of a few).}
\begin{equation}
b_{ij}(x) \rightarrow 
\left(\matrix{x-w & \ln x\cr
\ln x & w - x^{-1}}\right),
\quad 
w\equiv (1/4)!(-1/4)!
\label{eqn:bij5}
\end{equation}
The constant $w=(1/4)!(-1/4)!$ is obviously\footnote{because the factorial
function is logarithmically convex and $0!=1$}  just slightly larger than
unity, $w\simeq 1.111$.
The naive argument given in Section~\ref{sec:naive} maps
directly onto Equations~(\ref{eqn:bij4}) and (\ref{eqn:bij5}).
The prefactor $3\pi nD_*^3=3N_*$ is (3 times) the number of RR Lyrae stars
within the radius $D_*$ at which the astrometric and PL-relation errors
are equal.  The information content about $\pi_0$ is equivalent
to a naive integral outside this radius, $b_{11} \simeq N_*(x-1)/\epsilon^2$.
The reason that the Gaia precision constant $\kappa$ does not explicitly
enter this formula is that the volume element $(r^2)$ exactly cancels
the distance dependence of the inverse square of the errors $(\pi/\kappa)^2$.
Hence the amplitude of this essentially constant integral is set at $D_*$
where $\kappa/\pi = \epsilon\pi$.

The information content about the PL relation is equivalent to a naive
integral within most of the interior volume  
$b_{00} \simeq N_*(1-1/x)/\epsilon\kappa$.

Finally, the formal mathematical quantification of the ``circular
argument'' given in Section~\ref{sec:naive} is the correlation coefficient
$\rho$
\begin{equation}
\rho(x) = -{\ln x\over \sqrt{(x - w)(w-x^{-1})}}
\label{eqn:cc}
\end{equation}
For modest values of $x_\max$, $\rho$ is quite large.  For example,
$\rho(2.5,3,4)=-(0.92,0.91,0.88)$.  These high values degrade
the naive information content about $\pi_0$ by
\begin{equation}
[\sigma(\pi_0)]^2 = C_{00} = {\epsilon^2\over 3N_*}{b_{00}^{-1}\over 1-\rho^2}
\label{eqn:c00}
\end{equation}
where $C\equiv B^{-1} $ is the covariance matrix and $b_{00}\simeq x-w$

Before applying these equations to the problem of measuring $\pi_0$,
we must first account for the fact that Gaia precisions do not
further improve as the source gets brighter than G=12.  For
RR Lyrae stars, this corresponds to distance $D_\min=1.9\,\kpc$,
and so to 
\begin{equation}
x_\min = {D_\min/D_*} = 1.08\biggl({\epsilon\over 0.01}\biggr)^{-1/2}
\label{eqn:xmin}
\end{equation}
Then the formula for the inverse covariance matrix $B$ remains
valid provided one substitutes
\begin{equation}
b_{ij}\rightarrow b_{ij} - \Delta b_{ij}
\label{eqn:adjust}
\end{equation}

where
\begin{equation}
\Delta b_{ij}=\int_0^{x_\min}dy{y^{4-i-j}\over 1+y^4}-{y^{4-i-j}\over 1+y^2 x_\min^2}
\label{eqn:adjust2}
\end{equation}
In the relevant range of $x_\min$, this adjustment is quite small and
below the level of the errors made by various other approximations
in this treatment.  For example 
$\Delta b_{ij}(x_\min=1)=(0.014,0.020,0.020,0.028)$.

\section{{Numerical Estimates}
\label{sec:numerical}}
\begin{figure}
\centering
\includegraphics[width=85mm]{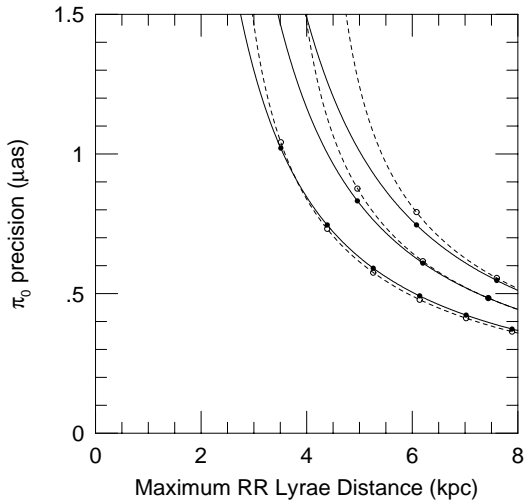}
\caption{Precision of Gaia estimate of $\sigma(\pi_0)$ based on
RR Lyrae star method, assuming all RR Lyrae stars are incorporated out
to a maximum distance indicated on the abscissa.  
Solid lines show result of numerical
integration assuming intrinsic distance scatter from the PL relation
of $\epsilon=(0.01,0.02,0.03)$ (bottom to top), while dashed lines
show the analytic approximation given by Equation~(\ref{eqn:bij5}).
Open and filled circles are at $x=(2.0,2.5,3.0,\ldots)$, showing
that the approximation becomes essentially exact for $x\geq 2.5$
\label{fig:prec}}
\end{figure}

To make numerical estimates of the precision that can be achieved, we first estimate $n_0=5.8\,\kpc^{-3}$ based on Hipparcos RR Lyrae stars
that are $V<11$ and that satisfy the \citet{layden96} ``Halo-3''
criteria (which were also adopted by \citealt{popow98}
and \citealt{gould98}).  The restrictive magnitude limit is to ensure 
completeness.  Of course, the so-called ``thick disk'' RR Lyrae stars
that do not satisfy the ``Halo-3'' criteria will also contribute
to the determination.  They are nevertheless excluded to be conservative
because their flattened three-space distribution implies that they
will contribute much less leverage at large distances compared to
halo RR Lyrae stars.

Figure~\ref{fig:prec} shows the estimated precision $\sigma(\pi_0)$
that can be achieved assuming that the RR Lyrae distance scatter
in a particular IR band is $\epsilon=0.01,$ 0.02, or 0.03, and
as a function of the upper distance limit $r_\max$ to which are RR Lyrae
measurements are essentially complete.  At, for example, $r_\max=5\,\kpc$,
these precisions are $\sigma(\pi_0)=(0.63,0.82,1.01)\,\muas$ for 
$\epsilon=(0.01,0.02,0.03)$.  This shows that the RR Lyrae star method
is comparable to the quasar method.

The solid lines in Figure~\ref{fig:prec} show the results of numerical
integration, i.e., using Equations~(\ref{eqn:bij4}), (\ref{eqn:adjust}),  
and (\ref{eqn:adjust2}), while the dashed lines show the results of
using the analytic approximation in Equation~(\ref{eqn:bij5}).  The
solid and open circles on these curves denote the evaluations at
$x=(2.0,2.5,3.0\ldots)$.  As predicted analytically in the text above,
the approximations are essentially perfect for $x\geq 2.5$.

Of course, RR Lyrae stars are not distributed uniformly around the Sun,
but the formalism developed here only requires that this be true
averaged over shells.  Even this assumption is not strictly valid,
but remains approximately valid for $r<8\,\kpc$, since the declining
density toward the Galactic anti-center is compensated by the increasing
density toward the Galactic center.  However, the approximation becomes
completely invalid for $r\gtrsim 8\,\kpc$ since at that point the
density is declining in all directions.  Hence, the fundamental limits
of the method are illustrated by the abscissa cut-off in Figure~\ref{fig:prec}.

\section{{RR Lyrae Distance Scale}
\label{sec:distance}}

It is also of interest to estimate how well the zero point of the PL
relation can be determined.  From algebraic manipulation of Equations~(\ref{eqn:bij4}) and  (\ref{eqn:bij5}), this is related to $\sigma(\pi_0)$ by
\begin{equation}
{\sigma(A)\over\sigma(\pi_0)} = 
\sqrt{{\epsilon\over\kappa}\,{x-w\over w-x^{-1}}}
= {D_*\over\au}\sqrt{x-w\over w-x^{-1}}
\label{eqn:sigmaA}
\end{equation}
Since, $\sigma(\pi_0)\sim {\cal O}(\muas)$, while  
$\au/D_*\sim {\cal O}(\mas)$, this implies that the zero point of
the RR Lyrae PL relation can be measured with precision of order
$10^{-3}$.  From RR Lyrae itself (and 3 other RRab stars), the absolute zero point is known to approximately $5\%$ \citep{benedict11} using trigonometric parallaxes from HST.  For RRc variables, the most precise values of the zero point come from the trigonometric parallax of RZ Cep \citep{benedict11} and the statistical parallax analysis from the CARRS survey \citep{kollmeier13}, although these values are in marginal tension.  As demonstrated above,  {\it Gaia} precision will be dramatically superior.

\section{{Conclusion}
\label{sec:conclude}}
The Gaia mission data promises to transform our understanding of the Milky Way.  In this work, we have shown that exploiting photometric parallax estimates for RRL within 8 kpc in conjunction with the precise IR P-L relation for these objects, one can measure the absolute parallax zero-point $\sigma(\pi_0)$ to precision of less than $0.5(\muas)$.  Not only is this extremely precise, but it is also comparable to, and completely independent of, measurements of this quantity from quasars.  We further show that once this is determined, one can refine the precision of the IR P-L zero point well beyond what is possible from photometric measurements alone.  We anticipate this independent method will be of immediate use to the astronomical community.


\acknowledgments
This work was supported by NSF grant AST-1516842.


\end{document}